\documentclass[a4paper]{jpconf}
\usepackage{graphicx}
\def\bea{\begin{eqnarray}}
\def\eea{\end{eqnarray}}
\begin{document}
\title{Impact of Nuclear Effects in the Measurement of Neutrino Oscillation Parameters}

\author{Davide Meloni}

\address{Dipartimento di Fisica "E. Amaldi", Universit\'a degli Studi Roma Tre, Via della Vasca Navale 84, 00146 Roma}

\ead{meloni@fis.uniroma3.it}

\begin{abstract}
In recent years the experimental study of neutrino oscillations has much contributed to our knowledge of particle physics by establishing non vanishing neutrino masses 
and by measuring or constraining the corresponding mixing angles. Within the domain of neutrino oscillations, the main goal of the next generation of facilities 
is the measurement of the mixing angle $\theta_{13}$ and the observation of leptonic CP violation, 
for which we have no hints at the moment. 
We discuss more in 
detail how various models of neutrino-nucleus cross section affect the forecasted precision 
measurement of $\theta_{13}$ and the CP violating phase $\delta$ \cite{FernandezMartinez:2010dm}.\\
\it Contribution to
NUFACT 11, XIIIth International Workshop on Neutrino Factories, Super beams and Beta
beams, 1-6 August 2011, CERN and University of Geneva
(Submitted to IOP conference series)
\end{abstract}

% \section{Introduction}
% In recent years the experimental study of neutrino oscillations has much contributed to our knowledge of particle physics by establishing non vanishing neutrino masses 
% and by measuring or constraining the corresponding mixing angles. Within the domain of neutrino oscillations, the main goal of the next generation of facilities,
% like  $\beta$-Beams~\cite{Zucchelli:sa},
% is the measurement of the mixing angle $\theta_{13}$ and the observation of leptonic CP violation, 
% for which we have no hints at the moment. 
% In this talk, we discuss more in 
% detail how various models of neutrino-nucleus cross section affect the forecasted precision 
% measurement of $\theta_{13}$ and the CP violating phase $\delta$ \cite{FernandezMartinez:2010dm}.

\section{Summary of the charged current neutrino-nucleus cross sections}
\label{summary}

In the quasi-elastic regime (QE), the doubly-differential cross section, in which a neutrino carrying initial four-momentum $k=(E_\nu,\bf k)$ scatters off a nuclear target to a 
state of four-momentum
$k^{'}=(E_\ell,\bf k^{'})$ can be written in Born approximation as follows:
\bea
\label{elem}
\frac{d^2\sigma}{d\Omega dE_\ell}=\frac{G_F^2\,V^2_{ud}}{16\,\pi^2}\,
\frac{|\bf k^{'}|}{|\bf k|}\,L_{\mu\nu}\, W_A^{\mu\nu} \ , 
\eea
where $G_F$ is the Fermi constant and $V_{ud}$ is the CKM matrix element coupling $u$
and $d$ quarks. The leptonic tensor is completely determined by lepton kinematics, whereas the nuclear tensor 
$W_A^{\mu\nu}$, containing all the information on strong interactions dynamics,
describes the response of the target nucleus,
% \bea
% \label{hadronictensor}
$W_A^{\mu\nu} =  \sum_X \,\langle 0 | {J_A^\mu}^\dagger | X \rangle \,
      \langle X | J_A^\nu | 0 \rangle \;\delta^{(4)}(p_0 + q - p_X),$ 
% \eea
where $|0\rangle$ and $|X\rangle$ are the initial and final hadronic states
carrying four momenta $p_0$ and $p_X$ and $J^\mu_A$ is the nuclear
electroweak current operator;
the sum includes all hadronic final states.
In the impulse approximation (IA) scheme,
the nuclear current can be written as a sum of one-body
currents, i.e. $J^\mu_A \rightarrow  \sum_i \, J^\mu_i$, while the final state reduces to the direct product
of the hadronic state produced at the weak vertex (with momentum ${\bf p^{'}}$)
and that describing the $(A-1)$-nucleon residual system, with momentum $\bf p_{\cal R}$:
$| X \rangle \to | i,{\bf p}^{'} \rangle \otimes | {\cal R}, \bf p_{\cal R} \rangle$. 

\subsection{The Spectral Function approach}

Following Ref. \cite{Benhar:2005dj}, the final expression of the hadronic tensor can be cast in the following form:
\bea
W_A^{\mu\nu}&=& \frac{1}{2}\int d^3p\,dE \,P({\bf
p},E)\frac{1}{4\,E_{|\bf p|}\,E_{|\bf p+q|}} \,  W^{\mu\nu}(\tilde p,\tilde
q) \ ,
\label{hadtensor}
\eea
where $E_{\bf p}=\sqrt{|{\bf p}|^2+m_N^2}$ and the function $P({\bf p},E)$ is the target {\it Spectral Function}, i.e.
 the probability distribution of finding a nucleon with momentum 
${\bf p}$ and removal energy $E$ in the target nucleus. The quantity $W^{\mu\nu}$ is the tensor describing
the weak interactions of the $i$-th nucleon in free space; the effect of
nuclear binding of the struck nucleon is accounted for by the
replacement $q=(\nu,{\bf q}) \to \tilde q=(\tilde \nu,{\bf q})$ 
with $\tilde \nu = E_{|\bf p+q|}-E_{|\bf p|}$. 
The second argument in the hadronic tensor is $\tilde p=(E_{|\bf p|},\bf p)$.
% The Spectral Functions for medium-heavy nuclei have been modeled using the
% Local Density Approximation (LDA) \cite{bffs}, in which the experimental
% information obtained from nucleon knock-out measurements is combined
% with the results of theoretical calculations of the nuclear matter
% $P({\bf p},E)$ at different densities.

\subsection{The Relativistic Fermi Gas}
The RFG \cite{Smith:1972xh} model, widely used in MonteCarlo simulations, provides the 
simplest form of the Spectral Function,
% \bea
% \label{fermigas}
$P_{RFGM}({\bf p},E)=\left(\frac{6\,\pi^2\,A}{p_F^3}\right)\,\theta(p_F-{\bf p})\,
\delta(E_{\bf p}-E_B+E),$
% \eea
where $p_F$ is the Fermi momentum and $E_B$ is the average binding energy,
introduced to account for nuclear binding. 
Thus, in this model $p_F$ and $E_B$ are two parameters that are {\it adjusted} to 
reproduce the experimental data. For oxygen, the analysis of electron scattering 
data yields $p_F=225$ MeV and $E_B=25$ MeV.

\subsection{The Relativistic Mean Field approach}
Within the Relativistic Mean Field approximation we refer to the model described in \cite{Martinez:2005xe}, 
where, like in the previous cases, the nuclear current is written as a sum of single-nucleon
currents. The wave functions for the target and the residual nuclei are described in
terms of an independent-particle model. Then, the transition matrix elements can be cast
in the following form:
\begin{equation}
J^{\mu}_{N}(\omega,\vec{q})=\int\/\/ d\vec{p}\/
\bar{\psi}_F
(\vec{p}+\vec{q}) \hat{J}^\mu_N(\omega,\vec{q}\/) \psi_B(\vec{p})\; ,
\label{nucc}
\end{equation}
where  $\psi_B$ and $\psi_F$ are the wave functions for initial bound
and  final outgoing  nucleons, respectively, and $\hat{J}^\mu_N$ is
the relativistic current operator. In particular, the relativistic bound-state wave functions 
(for both initial and outgoing nucleons) are obtained as a solution of the Dirac equation,
in the presence of the same relativistic nuclear mean field potential, derived from a Lagrangian
containing $\sigma$, $\omega$ and $\rho$ mesons. 

\subsection{The Random Phase Approximation}
The last model we want to take into account has been introduced in \cite{Martini:2009uj}, where the 
hadronic tensor is expressed in terms of the nuclear response functions 
treated in the  Random Phase Approximation (RPA). The response functions are related to the imaginary part of the 
corresponding full polarization propagators and the introduction of the RPA approximation means that 
the polarization propagators are the solutions of integral equations involving the bare propagators and 
the effective interaction between particle-hole excitations. Within this formalism, the authors of \cite{Martini:2009uj}
were able to show that multinucleon terms sizably increase the genuine charged current QE cross section 
in such a way to reproduce the MiniBooNE results \cite{AguilarArevalo:2010zc}.
The mechanism responsible for the enhancement that brings the theoretical cross section into agreement with the data
is multi nucleon knock out, leading to two particle-two hole (2p2h) nuclear final states. In the following, we will refer to
this ``generalized''  QE cross section as RPA-2p2h whereas we adopt the short RPA for the genuine QE cross section.

\subsection{Comparison of the cross sections}
\begin{figure}[h!]
\centering
\includegraphics[width=0.43\linewidth]{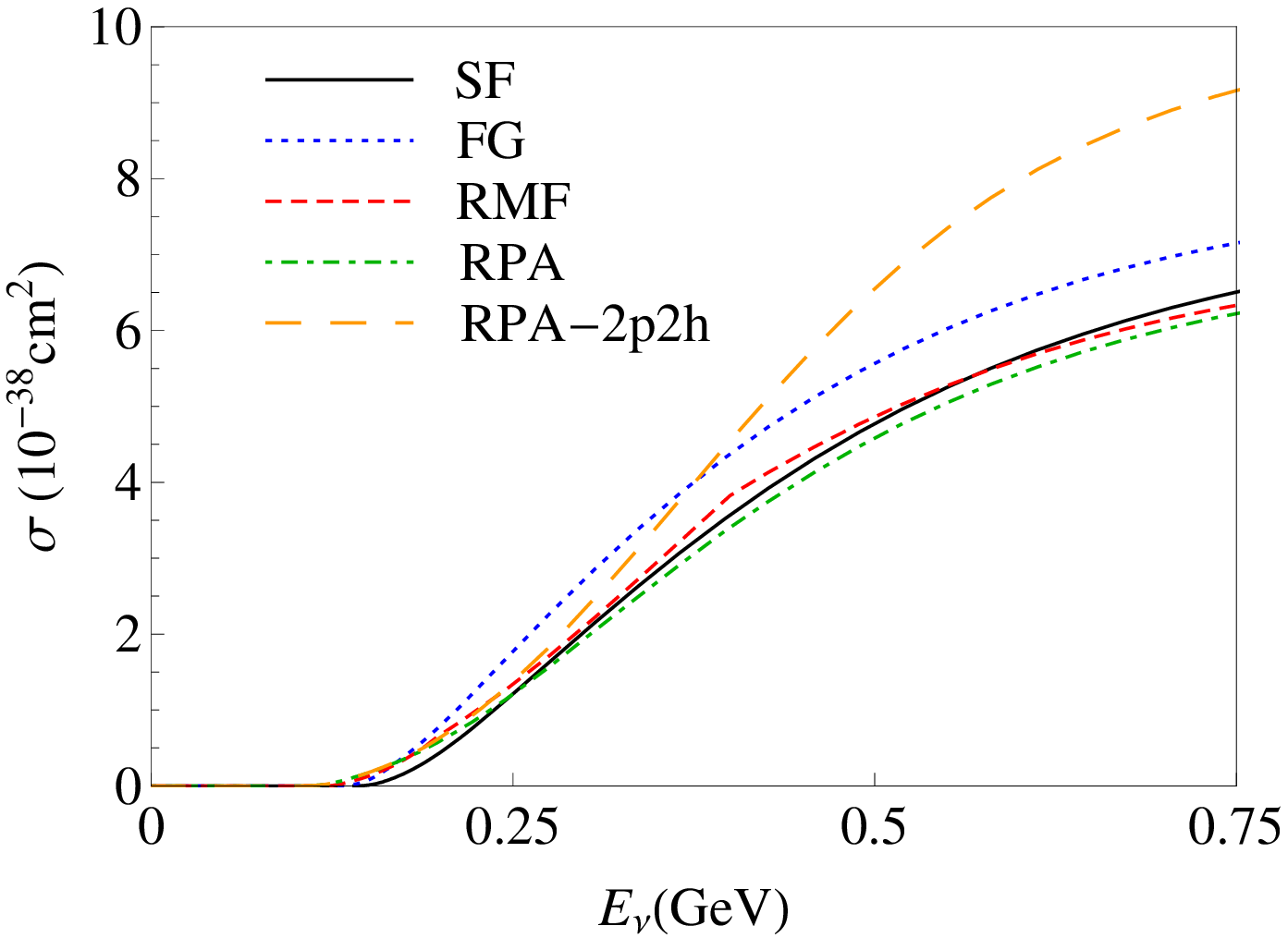}
\includegraphics[width=0.43\linewidth]{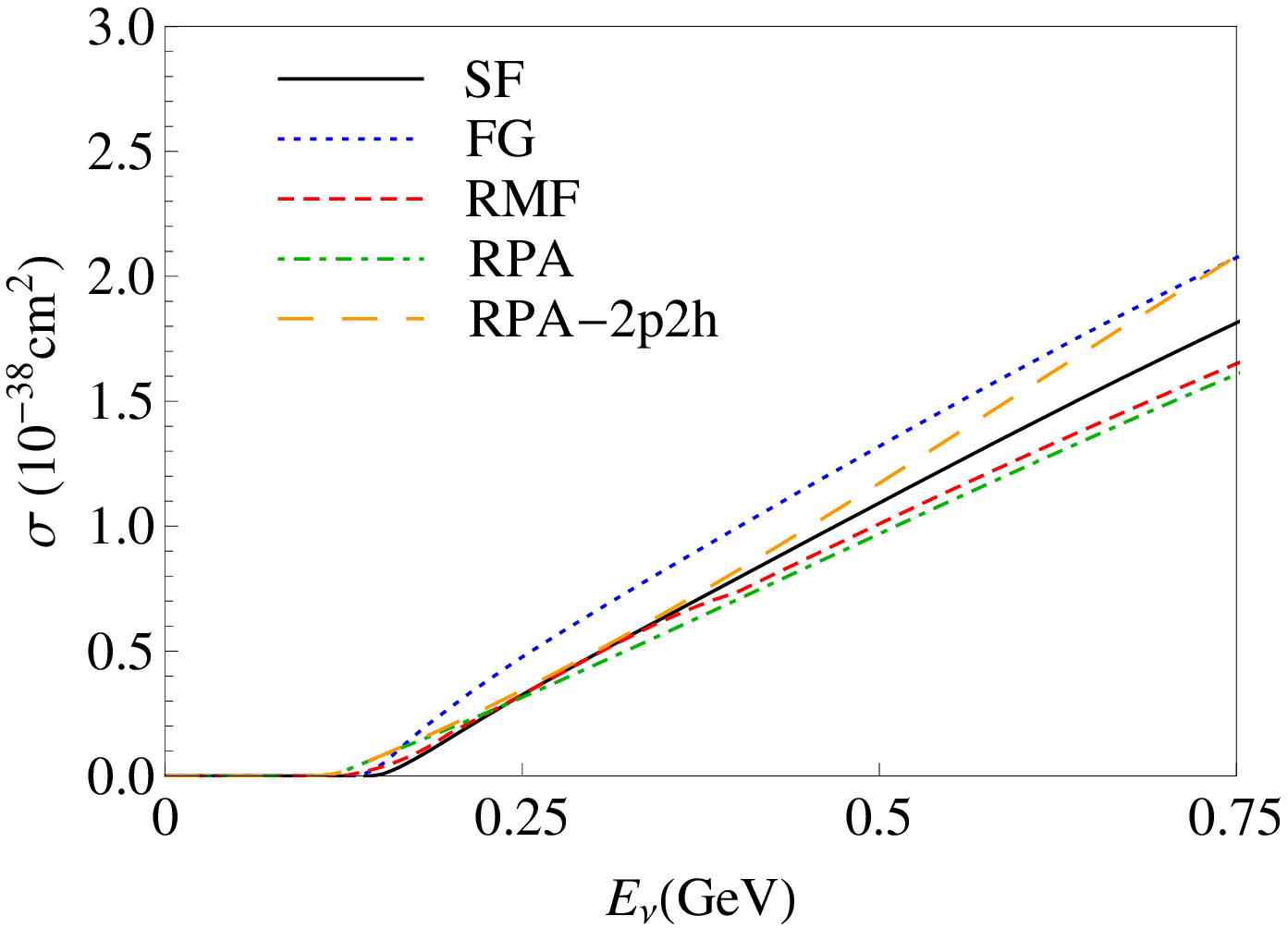}
\caption{\label{fig:xsect} \it Total charged current QE cross sections 
for the $\nu_\mu ^{16}O \to \mu^- X$ (left panel) and  $\bar \nu_\mu ^{16}O \to \mu^+ X$ (right panel) processes in the energy range $E_\nu\sim [0,0.75]$ GeV.}
\end{figure}

To summarize this section, we present in Fig.\ref{fig:xsect} a comparison of the five total QE cross sections 
for the $\nu_\mu \,^{16}O \to \mu^- X$ process (left panel) and $\bar \nu_\mu \,^{16}O \to \mu^+ X$ (right panel), in the energy range $E_\nu\sim [0,0.75]$ GeV. 
The curves have been computed using the dipole structure of the form factors and, in particular, a value of the axial mass
close to $m_A\sim 1$ GeV.
As it can be easily seen in the left panel, the RFG prediction sizably overestimates the SF, RMF and RPA results by roughly $15\%$, a fact that is well known 
to happen also for many other models with a 
more accurate description of the nuclear dynamics than the RFG approach. On the other hand, the inclusion of 2p2h contributions largely
enhances the QE cross section in the RPA approximation for energies above $\sim 0.5$ GeV, although it is smaller than the RFG at smaller energies.
For antineutrinos (right panel) the observed pattern is almost the same. 
 
\section{The impact on the ($\theta_{13}$-$\delta$) measurement}
\label{results}
\begin{figure}[h!]
\centering
\includegraphics[width=0.375\linewidth]{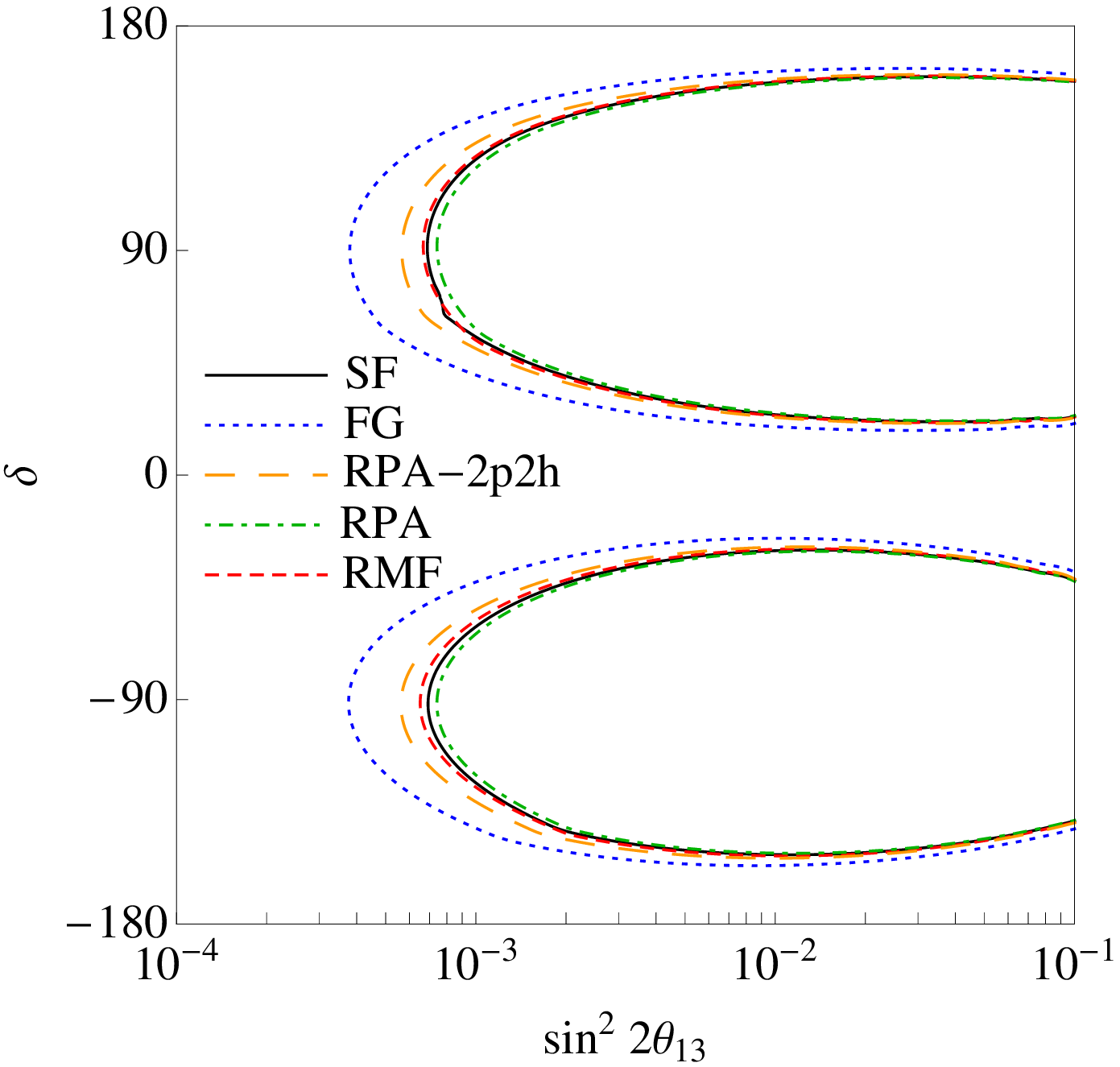}
\includegraphics[width=0.375\linewidth]{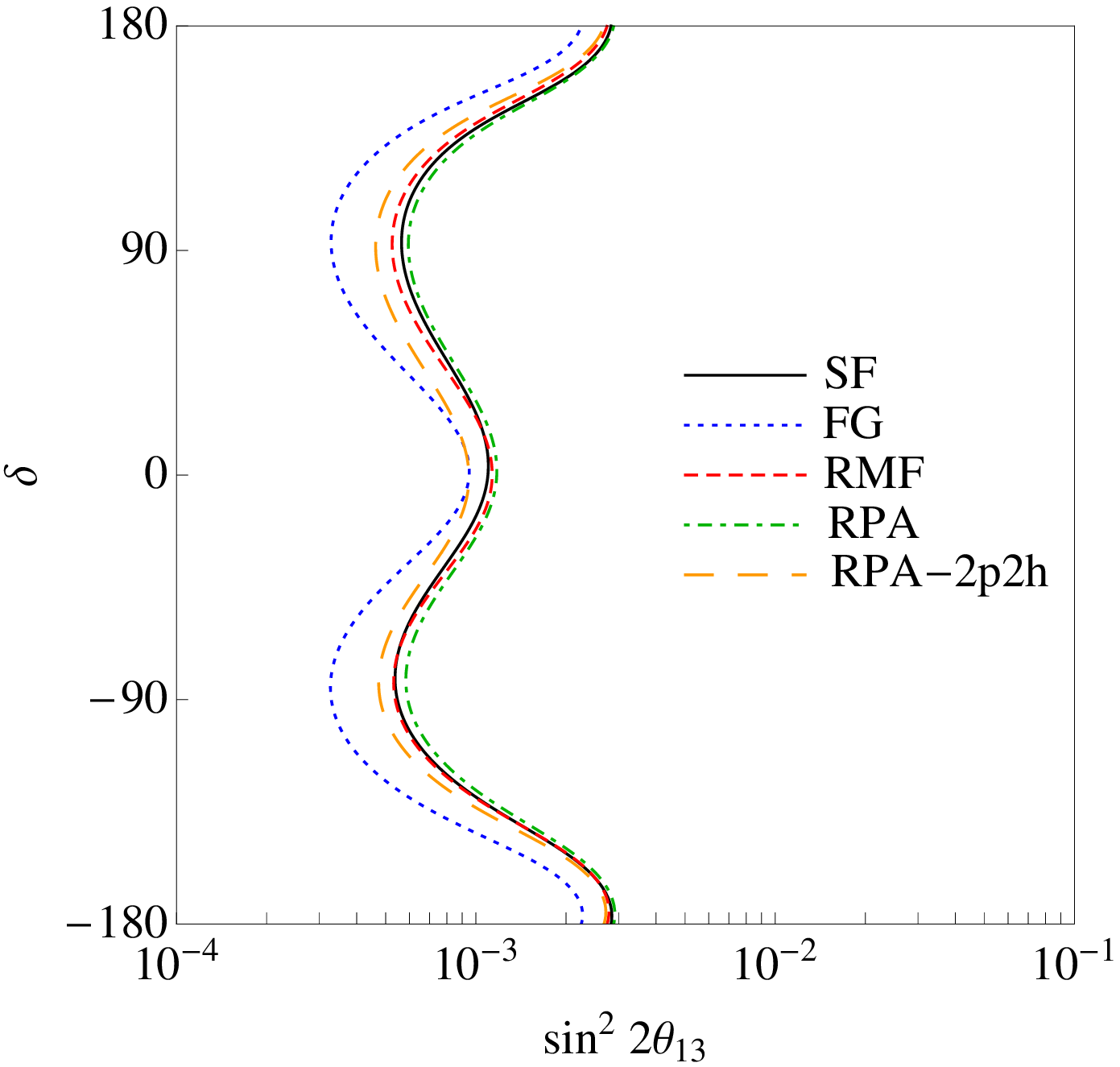}
\caption{\label{fig:disc} \it  Left panel: CP discovery potential in the $(\theta_{13},\delta_{CP})$ plane. 
Solid lines refer to the SF model, dotted lines to the RFG, short-dashed lines to the RMF, dot-dashed to RPA and long-dashed to RPA-2p2h.
Right panel: $\theta_{13}$ discovery potential.}
\end{figure}

To estimate  the impact of different models of the cross section
on the measurement of $\theta_{13}$ and leptonic CP violation we choose a $\gamma = 100$
$\beta$-Beam facility as a representative example because the neutrino flux from such a facility spans up to $\sim 0.7$ GeV with the peak around $0.3$ GeV and is 
thus mostly sensitive to the quasielastic region explored here. 
Details of the detector response can be found in Ref.~\cite{Burguet-Castell:2005pa}. 
 It is important to notice that we are only using the quasielastic contribution to the neutrino cross section depicted in Fig.\ref{fig:xsect}. 
As an illustration we have focused on the dependence on the nuclear model adopted of two different observables, namely the {\it CP and $\theta_{13}$ discovery  potentials}, 
defined as the values of 
the CP-violating phase $\delta_{CP}$ and $\theta_{13}$ for which respectively the hypothesis of CP conservation $\delta_{CP}=0, \pm \pi$ or $\theta_{13}=0$ can be 
excluded at 3$\sigma$ after marginalizing over all other 
parameters. The CP discovery  potential is shown in the left panel of Fig.\ref{fig:disc}, where we superimposed the results obtained using 
the RFG cross section and the SF, RMF and RPA calculations.
We clearly see that, for $\delta_{CP}\sim \pm 90^o$ (where the sensitivity is maximal) 
the RFG model gives a prediction which is around a factor 2 better than the SF, RMF and RPA models (which, as expected, behave almost in the same way) in $\sin^2 2\theta_{13}$ and around a 
$40\%$ better than the RPA-2p2h. 
This is not surprising because the $\beta$-Beam facility used in our simulations mainly probes energies smaller than $0.5$ GeV, 
where the RFG is still larger than any other model 
(see Fig.\ref{fig:xsect}). For the other points in the parameter space, the difference is less evident but still significant. 
The difference in the sensitivity to $\theta_{13}$ can be seen in the right panel of Fig. \ref{fig:disc} where        
we show the results in the $(\sin^2 2\theta_{13},\delta_{CP})$-plane. 
In this case the predictions of the  SF, RMF and RPA models differ by up to a factor of $\sim 60\%$ compared to the RFG model for $\delta_{CP}\sim \pm 90^o$ while the difference is less pronounced for $\delta_{CP}\sim 0^o$. The RPA-2p2h results are more similar to the RFG for $\delta_{CP}\sim 0^o$ and to the other models for $\delta_{CP}\sim 180^o$. This is also evident in the right panel where we show the CP fraction.

Finally, it is interesting to comment on the effect of using different nuclear models also in the simultaneous determination of $\theta_{13}$ and $\delta_{CP}$.
We have performed such an analysis for the input value $(\theta_{13},\delta_{CP})=(0.9^o,30^o)$ (not including degenerate solutions  coming from our ignorance of the  
octant of $\theta_{23}$  \cite{Meloni:2008bd} and in the neutrino mass ordering) finding that, using the RFG and RPA-2p2h models, we are able to reconstruct the 
true values of $\theta_{13}$ and $\delta_{CP}$ within reasonable uncertainties, whereas with the other models
we can only measure two distinct disconnected regions (the fake one around the value of $\theta_{13}$ and $\delta_{CP}\sim 180^o$), which worsen the global sensitivity on those parameters. 
In conclusion, we have analyzed the impact of five different theoretical models for the neutrino-nucleus charged current QE cross sections on the forecasted sensitivity of the 
future neutrino facilities to the parameters $\theta_{13}$ and $\delta_{CP}$. We have found that the sensitivities computed with the FG
are better, by up to a factor of 2, with respect to the other models.

\ack
We  acknowledge  MIUR (Italy), for financial support
under the contract PRIN08.


\end{document}